\documentclass[journal]{IEEEtran}
\IEEEoverridecommandlockouts

\usepackage{cite}
\usepackage{amsmath,amssymb,amsfonts}
\usepackage{graphicx}
\usepackage{textcomp}
\usepackage{xcolor}
\usepackage{booktabs}
\usepackage{multirow}
\usepackage{url}
\usepackage{subfig}
\usepackage{array}
\usepackage{balance}
\usepackage{stfloats}

\title{Measuring Cross-Jurisdictional Transfer of Medical Device Risk Concepts with Explainable AI}

\author{Yu~Han
and~Aaron~Ceross%
\thanks{Yu Han is with the Institute of Biomedical Engineering, Department of Engineering Science, University of Oxford, Oxford OX1 3PJ, United Kingdom (e-mail: yu.han@eng.ox.ac.uk).}%
\thanks{Aaron Ceross is with Birmingham Law School, University of Birmingham, Birmingham, United Kingdom.}%
}

\begin{document}

\maketitle

\begin{abstract}
Medical device regulators in the United States (FDA), China
(NMPA), and Europe (EU~MDR) all use the language of risk,
but classify devices through structurally different mechanisms.
Whether these apparently shared concepts carry transferable
classificatory signal across jurisdictions remains unclear. We
test this by reframing explainable AI as an empirical probe of
cross-jurisdictional regulatory overlap. Using 141,942 device
records, we derive seven EU~MDR risk factors, including
implantability, invasiveness, and duration of use, and evaluate
their contribution across a three-by-three transfer matrix.

Under a symmetric extraction pipeline designed to remove
jurisdiction-specific advantages, factor contribution is negligible
in all jurisdictions (all $\Delta$F1 $<$ 0.01), indicating that clean
cross-jurisdictional signal is at most marginal. Under
jurisdiction-specific pipelines, a modest gain appears only in the
EU~MDR-to-NMPA direction ($\Delta$F1 = +0.024), but
sensitivity analyses show that this effect is weak, context-dependent,
and partly confounded by extraction and representation choices.
Reverse-direction probes show strong asymmetry: FDA-derived
factors do not transfer meaningfully in any direction, and
NMPA-derived factors do not carry signal back to EU~MDR.
Zero-shot transfer further fails on EU~MDR Class~I
(F1 = 0.001), consistent with a mismatch between residual and
positional class definitions.

Overall, cross-jurisdictional transfer is sparse, asymmetric, and
weak. Shared regulatory vocabulary does not, under this
operationalisation, translate into strong portable classification
logic. The findings challenge a common assumption in
cross-jurisdictional regulatory AI and show how explainable AI
can be used to measure, rather than assume, regulatory overlap.
\end{abstract}

\begin{IEEEkeywords}
Medical device classification, cross-jurisdictional regulation, regulatory concept transfer, explainable AI, regulatory structure
\end{IEEEkeywords}

\section{Introduction}

\begin{figure*}[tbp]
\centering
\includegraphics[width=0.88\textwidth]{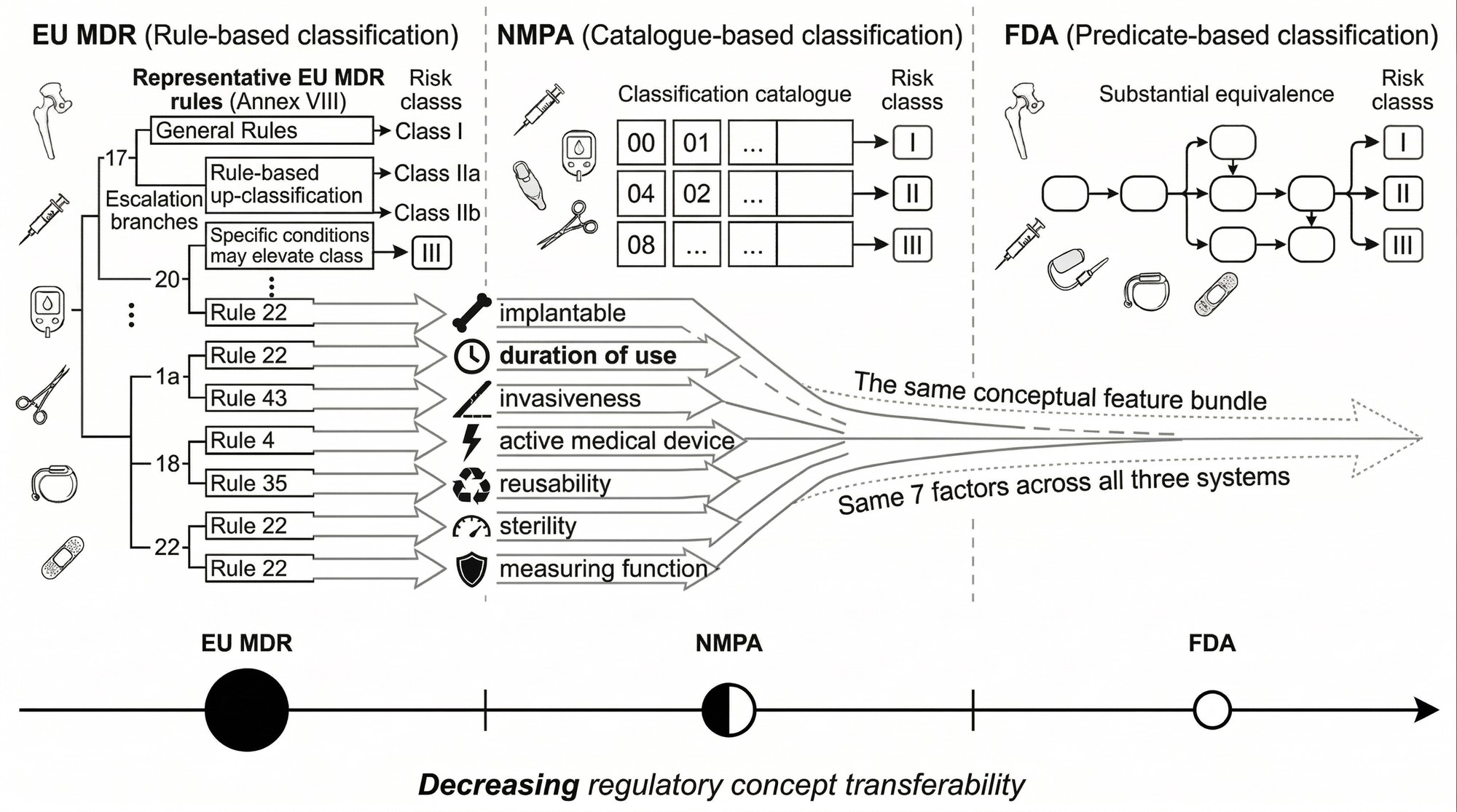}
\caption{Study overview. The EU~MDR classifies medical devices through 22 explicit rules in Annex~VIII, from which we extract seven regulatory risk factors (implantable, duration of use, invasiveness, active device, sterile, measuring function, and reusable). These factors serve as an empirical probe, testing whether risk concepts that carry classification signal in one regulatory system transfer meaningfully to structurally different ones such as NMPA's catalogue-based approach and FDA's predicate-based approach. The probe beam (centre) illustrates the observed gradient, with factor contribution highest within the originating rule-based system (a ceiling case reflecting circularity), dropping to modest levels in the catalogue-based system and falling to negligible in the predicate-based system.}
\label{fig:overview}
\end{figure*}

Medical device classification determines the regulatory pathway and post-market requirements that apply to a given product. All major jurisdictions assign devices to risk-based categories, but they do so through structurally different mechanisms \cite{mdr2017,fda2020,nmpa2021,fink2023international,chettri2024comparative}. The European Union Medical Device Regulation (EU~MDR) implements a rule-based system comprising 22 classification rules in Annex VIII, which assign devices to one of four risk classes (I, IIa, IIb, and III) based on characteristics such as invasiveness, duration of contact, and intended purpose. The United States Food and Drug Administration (FDA) employs a predicate-based system in which a device's classification depends substantially on whether a substantially equivalent device already exists on the market, with over 1,700 product codes encoding this regulatory history. China's National Medical Products Administration (NMPA) operates a catalogue-based system in which devices are matched to entries in a classification catalogue according to their product description and intended use.

These structural differences (Figure~\ref{fig:overview}) raise a fundamental question: how much regulatory structure do these systems actually share? Prior work has compared regulatory frameworks qualitatively \cite{muehlematter2021approval,muehlematter2023artificial,chen2024regulatory,reddy2025harmonization}, but no study has measured the degree of structural overlap empirically. Without such measurement, the growing number of AI systems proposed for regulatory decision support \cite{han2025ai,ceross2025toward,han2024china} must rely on untested assumptions about whether concepts like ``invasiveness'' or ``implantability'' carry the same classificatory weight across jurisdictions.

We test this question by treating classification as an empirical probe of regulatory structure. By deriving structured risk factors from one jurisdiction's classification logic and measuring whether those factors improve classification performance in other jurisdictions, we quantify the degree and directionality of cross-jurisdictional concept overlap. The resulting cross-transfer matrix, which tests three factor sources against three target jurisdictions, provides an empirical measure of regulatory structural distance across major device classification systems.

\subsection{Inferential Ladder}

The argument proceeds at three levels, from predictive gain to factor transferability to structural compatibility.
\begin{enumerate}
\item \textbf{Predictive gain} ($\Delta$F1) is the directly observed quantity. It measures the change in classification performance when regulatory factors are added to text features.
\item \textbf{Factor transferability} is the immediate interpretation. It asks whether a specific set of EU~MDR-derived factors carries classification signal in other jurisdictions.
\item \textbf{Structural compatibility} of regulatory systems is a higher-level inference. Establishing it would require considerably more evidence than this study provides, and we do not claim to do so.
\end{enumerate}

\noindent Accordingly, we treat predictive gain as evidence of transferability under this operationalisation, not as proof of structural compatibility. Observed $\Delta$F1 values support conclusions about factor transferability under the specific operationalisation tested; they do not directly prove or disprove structural compatibility between regulatory systems.

\subsection{Research Questions}

\textbf{RQ1.} Under one specific factorisation of EU~MDR Annex~VIII risk concepts, do these factors carry classification signal in other jurisdictions, and which risk classes benefit most?

\textbf{RQ2.} Under what structural conditions do cross-jurisdictional regulatory AI models fail, and what competing explanations remain live?

\subsection{Scope and Limitations}

Ground truth classification labels are derived from official database fields, specifically the NMPA UDI ``classification basis'' entries and FDA product code regulatory panel assignments, rather than from the reasoning chains of human reviewers. For FDA 510(k) devices, classification depends on predicate chains reflecting substantial equivalence, a mechanism that our factor-based approach does not capture.

The seven regulatory factors (implantable, duration of use exceeding 30 days, surgically invasive, active device, sterile, measuring function, and reusable) are derived from the classification rules in EU~MDR Annex VIII. EU~MDR is used as the source jurisdiction because, among the three systems examined here, it alone sets out its classification logic as explicit rules tied to device characteristics. By contrast, the FDA relies primarily on substantial equivalence to existing products, and NMPA assigns devices through catalogue matching based on product descriptions. Neither system provides an explicit rule structure from which comparable characteristic-based factors can be extracted directly.

For that reason, Annex VIII is the only available basis for constructing a systematically defined set of regulatory factors that can then be tested across jurisdictions. The analysis therefore examines transfer in one direction only, from the jurisdiction with codified characteristic-based rules to the two jurisdictions whose classificatory logic is less explicit. For EU~MDR devices, the factors are derived from the classification rule recorded in the database through a rule-to-factor mapping, supplemented by multilingual keyword extraction from product names. Because this procedure partially encodes the rules themselves, the EU~MDR factors contain a degree of circularity, which we address in Section~\ref{sec:circularity}.

The EU~MDR uses a four-class system (I, IIa, IIb, and III), while the FDA and NMPA each use three-class systems (I, II, and III). For the unified cross-regime model, we compress EU~MDR to three classes by merging IIa and IIb into a single Class~II category. To disentangle this compression artefact from other sources of cross-jurisdictional failure, we additionally evaluate EU~MDR under its native four-class scheme (Supplementary Table~S4).

\subsection{Related Work}

This study sits at the intersection of explainable AI, regulatory AI, and comparative medical device regulation, but its contribution is not simply to combine these literatures. We address a specific gap left by all three, as to whether apparently shared regulatory risk concepts travel across jurisdictions as operative classificatory signals rather than as surface-level vocabulary.

Within explainable AI, feature-attribution methods such as SHAP \cite{lundberg2017unified}, LIME \cite{ribeiro2016why}, and Integrated Gradients \cite{sundararajan2017axiomatic} identify influential inputs, but they do not by themselves test whether a model has captured semantically meaningful regulatory abstractions. This limitation matters in legal and regulatory settings, where the relevant structure is often organised around mid-level concepts (e.g., invasiveness, implantability, or duration of use) rather than around individual lexical features. Concept-based methods, including TCAV \cite{kim2018interpretability}, concept bottleneck models \cite{koh2020concept}, and completeness-oriented concept explanations \cite{yeh2020completeness}, are therefore closer to the object of analysis in this paper. We draw on this line of work not to explain a model to end users in the usual sense, but to test whether concept bundles derived from one regulatory regime retain classificatory value in another. We also borrow the perturbational logic used in counterfactual explanation research \cite{wachter2018counterfactual,mothilal2020explaining}, but use it more modestly as a diagnostic tool to investigate reclassification barriers rather than as a claim about causal counterfactuals.

In medical AI, explainability research has focused primarily on clinical decision support \cite{tjoa2020survey,amann2020explainability,bienefeld2023solving}, with a growing critique that many XAI techniques do not deliver the transparency or accountability they promise in practice \cite{ghassemi2021false}. Related work has begun to document transparency deficits in regulatory-authorised AI devices themselves \cite{shick2024transparency,muralidharan2024reporting,chouffani2024not}.Our paper addresses a different question: whether formalised risk concepts extracted from one jurisdiction retain classificatory value in another. We use the model as an empirical probe of regulatory structure to test the alignment between those concepts and cross-jurisdictional classification outcomes.

A separate body of work applies machine learning and language models to legal and regulatory tasks, including medical device classification and regulatory text processing \cite{han2025ai,ceross2025toward,chalkidis2020legalbert}. That literature has shown that specialized models can achieve useful within-domain performance, but it has generally treated regulatory labels as prediction targets rather than as evidence about the structure of the regimes that generate them. In particular, prior classification studies ask whether models can reproduce decisions within a jurisdiction; they do not test whether the concepts encoded in one regime remain informative when transferred to another. That distinction is central here. Instead of focusing on whether AI can classify devices accurately, our study examines whether cross-jurisdictional regulatory AI rests on a defensible assumption of conceptual portability.

Comparative work on medical device regulation has analysed differences and convergence points in the FDA, EU MDR, and NMPA systems \cite{muehlematter2021approval,muehlematter2023artificial,chen2024regulatory,han2024china}, while harmonisation efforts such as IMDRF \cite{imdrf2021samd} and recent reviews \cite{reddy2025harmonization} have articulated frameworks for cross-border alignment. The EU AI Act adds another layer of complexity for high-risk AI-enabled devices \cite{aboy2024navigating}. However, this literature remains largely doctrinal, descriptive, or policy-oriented. It identifies similarities and differences in regulatory architecture, but it does not test whether shared risk vocabulary corresponds to shared classificatory logic. That is the gap this paper addresses. By treating explainable AI as an empirical probe rather than as an end-user transparency tool, we test whether regulatory concepts derived from EU MDR carry measurable classification signal in NMPA and FDA data, and thereby assess the extent and asymmetry of cross-jurisdictional concept transfer.
\section{Methodology}

\subsection{Overview}

The study employs two methodological components, each corresponding to one research question. For RQ1, we conduct a controlled factor ablation study. For each jurisdiction independently, Random Forest classifiers are trained under three conditions: using both text features and regulatory factors, using text features alone, and using regulatory factors alone. Classification performance is evaluated with per-class F1 scores from five-fold stratified cross-validation. To assess the reliability of the factor contribution ($\Delta$F1), we repeat the five-fold cross-validation across 10 random seeds. For each seed, the mean $\Delta$F1 is calculated in five folds, yielding 10 summary statistics of the seed-level. We report the mean and 95\% confidence interval computed from these 10 seed-level means, treating each full cross-validation run as the independent unit of replication. We foreground effect size and uncertainty rather than null-hypothesis significance testing, as the five-fold design provides limited inferential granularity for $p$-values. All Random Forest classifiers use inverse class frequency weighting to address class imbalance (e.g., EU~MDR Class~III constitutes only 1.9\% of devices). This design isolates how much of each jurisdiction's classification logic is captured by EU~MDR-derived regulatory factors compared to text features. For RQ2, we use two diagnostic approaches. We first conduct zero-shot transfer experiments, evaluating a model trained on FDA and NMPA data on EU~MDR data without any EU~MDR training examples. We then apply a greedy perturbation-based reclassification search to identify the conditions under which valid reclassification paths cannot be found.

\subsection{Perturbation-Based Reclassification Search}

For each sampled device, we probe the trained model's decision boundaries by searching for factor perturbations that change its predicted class. This analysis characterises the classifier's learned behaviour rather than the regulatory systems themselves; a perturbation that fails to reclassify a device in the model does not imply that the regulatory system lacks a valid reclassification path. Given a device with feature vector $\mathbf{x}$ classified as class $c$ by the trained Random Forest, we seek a minimal modification $\mathbf{x}'$ such that the model predicts a different target class $c' \neq c$. The search proceeds as follows: (1) for each of the seven regulatory factors, we enumerate all valid alternative values (e.g., invasiveness $\in \{0, 1, 2\}$); (2) for each candidate perturbation, we query the trained model for its predicted class; (3) among all single-factor perturbations that achieve the target class $c'$, we select the one requiring the smallest change in factor value (i.e., a change from 0 to 1 is preferred over 0 to 2); (4) if no single-factor perturbation succeeds, we attempt all two-factor combinations, then three-factor combinations, up to a maximum of four simultaneous changes. A reclassification attempt is deemed successful if a perturbation achieving the target class is found within this search budget, and unsuccessful otherwise. We sample 500 devices per jurisdiction (stratified by class) and attempt reclassification to each alternative class, yielding up to 1,000 attempts per jurisdiction.

\subsection{Regulatory Factor Extraction}

Seven regulatory factors are extracted from device records using jurisdiction-specific methods. For NMPA devices, factors are inferred from the two-digit classification code prefix in the UDI record. For example, prefixes 03, 06, and 13 indicate implantable devices, while prefixes 08, 16, and 22 indicate devices with a measuring or diagnostic function. These code-based assignments are supplemented by keyword matching against the Chinese-language product description field.

For FDA devices, factors are extracted through keyword matching against the device name and summary fields. Terms such as ``implant,'' ``prosthesis,'' and ``stent'' indicate implantable devices, while terms such as ``catheter,'' ``surgical,'' and ``needle'' indicate surgically invasive devices.

For EU~MDR devices, a different approach is necessary because EUDAMED records contain limited textual information, averaging only 36 characters per product description. Factors are therefore derived primarily from the MDR Annex VIII classification rule recorded in the database. Each rule corresponds to a characteristic factor profile. For instance, Rule~8, which covers long-term surgically invasive or implantable devices, is coded as implantable = 1 and invasiveness = 2. Rule~1, which covers non-invasive devices, is coded with all risk factors set to 0 and reusable set to 1. We supplement this rule-based derivation with keyword matching on product names in English, German, and French.

The EU~MDR approach derives factors from the classification rule itself, which introduces circularity because the factors partially encode the classification outcome. This circularity is addressed in Section~\ref{sec:circularity}.

\subsection{Feature Representation}

Each device is represented by a 49-dimensional feature vector consisting of seven regulatory factors and 42 text-derived features. The text features are computed by applying term frequency-inverse document frequency (TF-IDF) weighting with character n-grams (1--2 characters, word-boundary aware) to device descriptions, limited to the 500 most frequent features, followed by dimensionality reduction to 42 components through truncated singular value decomposition (SVD). We chose a character n-gram analyser because it is language-agnostic and can operate across scripts without language-specific tokenisation. This is necessary because NMPA descriptions are in Chinese, FDA descriptions are in English, and EU~MDR descriptions span several European languages.

An important limitation of this representation is that TF-IDF with character n-grams produces disjoint feature spaces across different writing systems. In the per-jurisdiction experiments (Section~\ref{sec:leakage_audit}), this is benign because each model operates within a single language domain. In the unified and zero-shot experiments, however, the SVD components effectively encode language identity rather than shared semantic content, rendering the text channel non-functional across jurisdictions. To address this limitation, we conduct sensitivity analyses with multilingual sentence embeddings that operate in a shared cross-lingual semantic space (Section~\ref{sec:multilingual}).

\section{Experimental Setup}

\subsection{Dataset}

\begin{table}[t]
\centering
\caption{Dataset composition by jurisdiction and risk class}
\label{tab:dataset}
\begin{tabular}{lrrrr}
\toprule
\textbf{Jurisdiction} & \textbf{$n$} & \textbf{Class I} & \textbf{Class II} & \textbf{Class III} \\
\midrule
FDA (US) & 53,745 & 1,373 & 28,302 & 24,070 \\
NMPA (China) & 65,703 & 1,309 & 37,564 & 26,830 \\
EU MDR (EU) & 22,494 & 9,819 & 12,255 & 420 \\
\midrule
\textbf{Total} & \textbf{141,942} & & & \\
\bottomrule
\end{tabular}
\end{table}

The dataset comprises 141,942 medical device records drawn from three sources (Table~\ref{tab:dataset}). FDA data includes 29,681 510(k) premarket notification clearances and 24,064 premarket approval records, obtained through the openFDA API. NMPA data consists of 65,703 records from the UDI database maintained by China's National Medical Products Administration. EU~MDR data includes 22,494 records from the EUDAMED database, spanning four native risk classes: Class~I (9,819 devices), Class~IIa (9,903), Class~IIb (2,352), and Class~III (420). For the unified cross-regime model, Class~IIa and Class~IIb are merged into a single Class~II category.

\subsection{Experiments}

For RQ1, we conduct per-jurisdiction Random Forest ablation experiments comparing three feature conditions with per-class F1 scores and 95\% confidence intervals. We additionally audit the quality of factor extraction across jurisdictions, evaluate EU~MDR under its native four-class scheme, and analyse concept-to-class correlations per jurisdiction.

For RQ2, we report the unified model's per-jurisdiction performance breakdown, conduct zero-shot transfer from FDA and NMPA to EU~MDR, and perform perturbation-based reclassification failure analysis disaggregated by jurisdiction, class transition direction, and factor complexity.

\section{Results}

\subsection{RQ1: Regulatory Concept Transfer}

\subsubsection{Factor Extraction Audit}
\label{sec:manual_annotation}

The three extraction pipelines differ in epistemic status. The EU~MDR factors are derived from the recorded classification rules, the NMPA factors from the prefixes of the classification code, and the FDA factors from the matching of text keywords. Table~\ref{tab:extraction_audit} summarises coverage, validation against manual annotation (150 devices per jurisdiction, 450 total), and inter-rater reliability (IRR).

\begin{table}[t]
\centering
\caption{Factor extraction audit. (A)~Pipeline validation against manual annotation; undeterminable rate = proportion of factor--device pairs where text was insufficient for annotation. (B)~Inter-rater reliability (post-adjudication, $n$=90 devices). Per-factor IRR detail in Supplementary Table~S1.}
\label{tab:extraction_audit}
\begin{tabular}{lccc}
\toprule
\multicolumn{4}{l}{\textbf{(A) Per-jurisdiction extraction quality}} \\
\midrule
\textbf{Metric} & \textbf{EU~MDR} & \textbf{NMPA} & \textbf{FDA} \\
\midrule
Undeterminable rate$^\dagger$ & 51.7\% & 15.3\% & 19.8\% \\
Pipeline precision & 0.31 & 0.51 & 0.66 \\
Pipeline recall & 0.65 & 0.85 & 0.70 \\
Pipeline F1 & 0.42 & 0.64 & 0.68 \\
\midrule
\multicolumn{4}{l}{\textbf{(B) Inter-rater reliability (pooled, $n$=90 devices)}} \\
\midrule
Mean Cohen's $\kappa$ (7 factors) & \multicolumn{3}{c}{0.781 \quad (range: 0.577--0.966)} \\
Mean \% agreement & \multicolumn{3}{c}{91.9\% \quad (range: 88.2--98.3\%)} \\
\bottomrule
\multicolumn{4}{p{\columnwidth}}{\footnotesize $^\dagger$EU~MDR EUDAMED records average 36 characters. Coverage is adequate for sterile, duration of use, and reusable across all jurisdictions; limited for implantable, active device, and measuring function in EU~MDR only.} \\
\end{tabular}
\end{table}

EU~MDR shows a substantially higher undeterminable rate than the other jurisdictions. In total, 51.7\% of EU~MDR factor--device pairs could not be annotated from the available text, compared with 15.3\% for NMPA and 19.8\% for FDA. This difference is likely due to the extreme brevity of EUDAMED records. Pipeline F1 shows the opposite pattern from factor contribution. EU~MDR, which shows the largest factor-related classification gain, has the lowest extraction quality. This likely reflects the fact that EU~MDR factors are derived directly from classification rules rather than from keyword matching. Inter-rater reliability was substantial overall (mean $\kappa$ = 0.781 \cite{landis1977measurement}). Agreement was lowest for active device ($\kappa$ = 0.577), likely because $\kappa$ is attenuated under skewed marginals (10--18% prevalence)

\subsubsection{Primary Estimand: Symmetric Extraction Pipeline}
\label{sec:symmetric_pipeline}

To obtain the cleanest estimate of cross-jurisdictional factor contribution, we first report results under a symmetric pipeline that removes all jurisdiction-specific extraction advantages. This pipeline re-extracts all seven factors using a uniform text-only keyword matching method across all three jurisdictions, deliberately discarding the EU~MDR rule-to-factor mapping and the NMPA code prefix heuristics. The resulting factors are noisier but epistemically comparable across jurisdictions, making this the appropriate primary estimand for cross-jurisdictional comparison.

\begin{table*}[t]
\centering
\caption{Master transfer results. Symmetric $\Delta$F1 is the primary estimand (uniform keyword extraction, no pipeline advantages). Jurisdiction-specific $\Delta$F1 is the upper bound (includes pipeline asymmetry). 95\% CI on jurisdiction-specific $\Delta$F1 from 10 seed-level means (each seed = one five-fold CV run). EU~MDR is segregated as a source-jurisdiction ceiling (see Section~\ref{sec:circularity}); its $\Delta$F1 reflects circular rule derivation and is not commensurable with the cross-jurisdictional NMPA and FDA results.}
\label{tab:master_results}
\begin{tabular}{lccccl}
\toprule
\textbf{Jurisdiction} & \textbf{F1 (text+factors)} & \textbf{F1 (text-only)} & \textbf{$\Delta$F1 (symmetric)} & \textbf{$\Delta$F1 (jur.-specific) [95\% CI]} & \textbf{Interpretation} \\
\midrule
EU~MDR$^\dagger$ ($n$=22,494) & 0.992$\pm$0.004 & 0.847$\pm$0.017 & +0.006 & +0.192 [+0.187, +0.197] & Source-jurisdiction ceiling \\
NMPA ($n$=65,703)             & 0.843$\pm$0.007 & 0.822$\pm$0.008 & +0.004 & +0.024 [+0.023, +0.024] & Modest cross-jurisdictional gain \\
FDA ($n$=53,745)              & 0.898$\pm$0.007 & 0.894$\pm$0.007 & +0.003 & +0.006 [+0.005, +0.007] & Negligible \\
\bottomrule
\multicolumn{6}{l}{\footnotesize $^\dagger$EU~MDR: factors derived from classification rules; jurisdiction-specific/symmetric ratio = 32-fold (NMPA: 6-fold; FDA: 2-fold).} \\
\end{tabular}
\end{table*}

\begin{figure*}[tbp]
\centering
\includegraphics[width=0.95\textwidth]{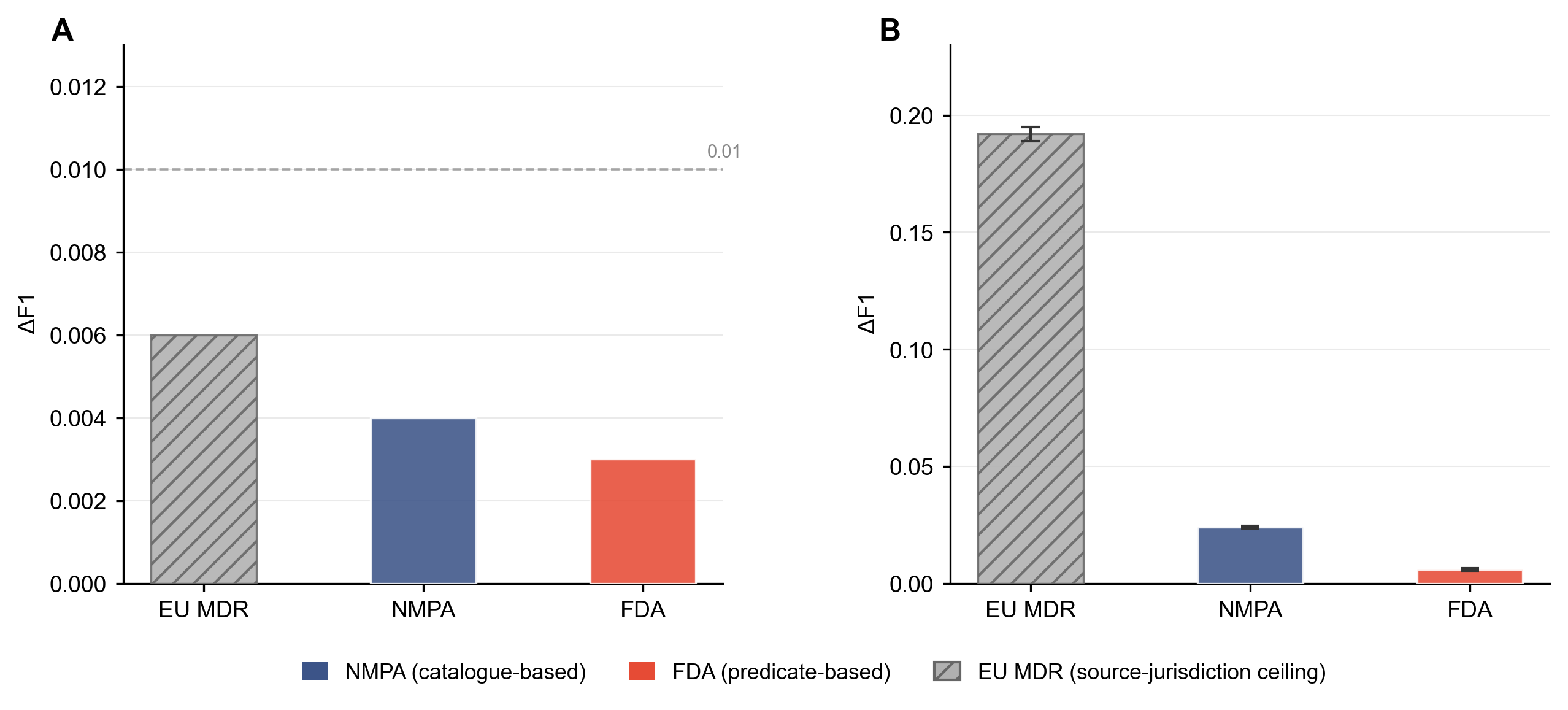}
\caption{Factor contribution ($\Delta$F1) under two estimands (EU~MDR hatched grey = source-jurisdiction ceiling, not commensurable with cross-jurisdictional results). (A)~Symmetric pipeline (primary estimand): all jurisdictions $\Delta$F1 $<$ 0.01, confirming the clean cross-jurisdictional signal is at most marginal. (B)~Jurisdiction-specific pipeline (upper bound): NMPA $+$0.024 [95\%~CI $+$0.023, $+$0.024]; FDA $+$0.006 [95\%~CI $+$0.005, $+$0.007]; EU~MDR ceiling $+$0.192 (circular, excluded from cross-jurisdictional inference). Error bars = 95\% CI from 10 seed-level means.}
\label{fig:headline}
\end{figure*}

Table~\ref{tab:master_results} presents both estimands side by side. Under the symmetric pipeline, factor contribution is near zero across all three jurisdictions ($\Delta$F1 $<$ 0.01), establishing that \textbf{the clean cross-jurisdictional signal, freed from pipeline-specific advantages, is at most marginal}. A nominal ordering across jurisdictions (EU MDR, NMPA, and FDA) is observed, but the corresponding effect sizes are weak and statistically indistinguishable from noise. However, the effect sizes are small and statistically indistinguishable from noise. Under the jurisdiction-specific pipeline, the NMPA gain rises to +0.024, approximately four times the FDA gain, but the six-fold ratio between the two estimands indicates that most of this gain is attributable to the extraction method rather than to genuine concept overlap. The FDA result is near-identical under both pipelines, consistent with predicate-based classification whose signal is already encoded in text features. Figure~\ref{fig:headline} visualises both estimands side by side; EU~MDR is visually segregated as a ceiling case in both panels.

\paragraph{Per-jurisdiction findings.} The NMPA gain (+0.024, 95\% CI [+0.023, +0.024]) is the only non-trivial cross-jurisdictional result and is concentrated in higher-risk classes (Class~III benefits most). The FDA gain (+0.006) is statistically non-zero but negligible in magnitude, consistent with predicate-based classification whose signal is already encoded in text. The EU~MDR ceiling result (factors-only F1 = 0.900, indicating substantial rule reconstruction) is expected given circular derivation and is not comparable to NMPA and FDA. Across all jurisdictions, factors-only fails for Class~I devices (F1 = 0.014--0.146), confirming that Class~I is defined by the \emph{absence} of risk factors.

\subsubsection{Circularity in EU~MDR Factor Derivation}
\label{sec:circularity}

For EU~MDR, the factors are derived from the classification rule recorded for each device in the database. They therefore partially encode the classification outcome itself. The high factors-only performance for EU~MDR (macro-F1 = 0.900) shows that these factors recover a substantial part of the rule structure.

This does not undermine the cross-jurisdictional analysis, but it does change how the EU~MDR result should be read. In the NMPA and FDA analyses, the same seven factors are extracted without this circularity: for NMPA from product descriptions and classification-code prefixes, and for FDA from device names and summary keywords. The EU~MDR result is therefore best treated as a source-jurisdiction ceiling, that is, the upper bound on factor utility when the factors encode the underlying classification logic. It should not be interpreted on the same footing as the NMPA and FDA transfer results (see also Section~\ref{sec:unified_ablation}).

Class compression has a minimal effect on EU~MDR performance (macro-F1: 0.804 three-class vs.\ 0.804 native four-class; Supplementary Table~S4). Concept-to-class associations (Cram\'{e}r's $V$) show that implantable is the strongest predictor across all three jurisdictions, but $V$ magnitudes are not directly comparable because EU~MDR has only 1.9\% Class~III devices; per-factor details are in Supplementary Figure~S8.

\subsection{RQ2: Failure Structure}

\subsubsection{Zero-Shot Transfer and Class~I Collapse}

\begin{figure}[tbp]
\centering
\includegraphics[width=\columnwidth]{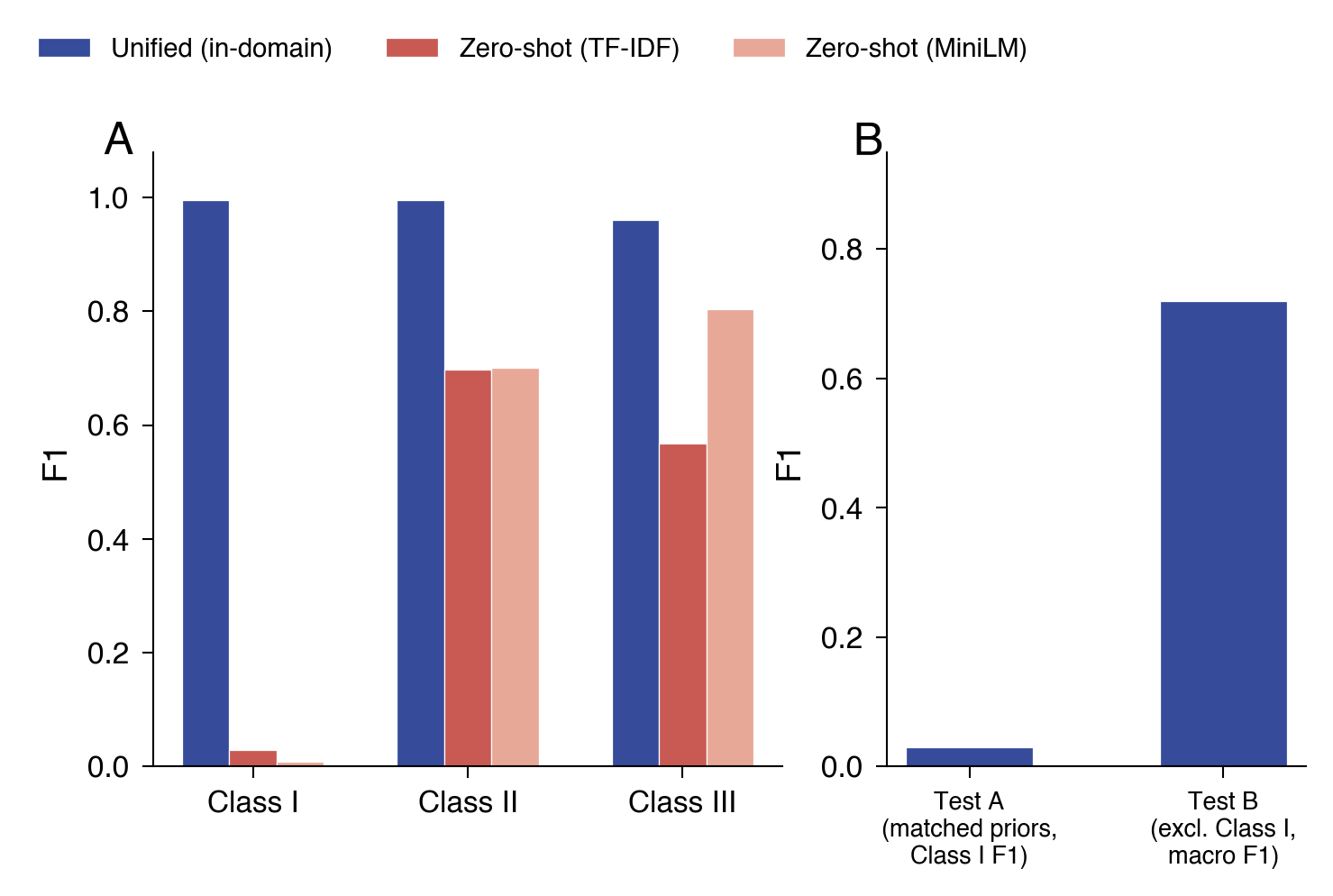}
\caption{EU~MDR zero-shot transfer failure. (A)~Classwise F1 under the unified model (EU~MDR in-domain) vs.\ zero-shot transfer (train FDA+NMPA, test EU~MDR) under the multilingual sensitivity protocol (Section~\ref{sec:multilingual}): Class~I collapses to near zero under both encoders (TF-IDF: 0.029; MiniLM: 0.009); under the main analysis protocol, the collapse is even more severe (Class~I F1 = 0.001, Table~\ref{tab:master_results}). (B)~Mechanism tests: Test~A (matched class priors, undersampling Class~I to 2.5\%) leaves Class~I F1 at 0.000, ruling out prior mismatch; Test~B (excluding Class~I) yields macro-F1 = 0.745, confirming Class~I collapse is the dominant degradation source.}
\label{fig:zero_shot_failure}
\end{figure}

The unified model achieves F1-macro of 0.984 for EU~MDR (high due to circular factor derivation), 0.805 for NMPA, and 0.864 for FDA; per-class details are in Supplementary Table~S5. Training on FDA and NMPA data and evaluating on EU~MDR without any EU~MDR training examples yields a macro-F1 of 0.384, with Class~I F1 falling to 0.001 (Figure~\ref{fig:zero_shot_failure}). Class~I constitutes the largest EU~MDR category with 9,819 devices (43.6\%), yet the model is effectively unable to identify them.

The failure likely arises because EU MDR Class I is residual, while FDA and NMPA Class I is positional, as devices that do not trigger any higher-class classification rule, whereas the FDA and NMPA define Class~I positionally through specific product codes or catalogue entries. However, the study design does not isolate this as the sole or dominant mechanism. There are competing explanations, which remain possible:
\begin{itemize}
\item \textbf{Class-prior mismatch}: EU~MDR has 43.6\% Class~I devices versus $\sim$2.5\% in FDA and NMPA, creating a severe distributional shift.
\item \textbf{Extraction asymmetry}: EU~MDR factors are derived differently from FDA and NMPA factors.
\item \textbf{Text domain shift}: EUDAMED's 36-character average descriptions differ substantially from FDA and NMPA text.
\item \textbf{Cross-lingual text barrier}: TF-IDF features computed on mixed Chinese (NMPA), English (FDA), and multilingual European (EU~MDR) text create language-partitioned feature spaces with near-zero cross-lingual information transfer, meaning the model relies almost entirely on the seven regulatory factors for cross-jurisdictional signal (Section~\ref{sec:multilingual}).
\item \textbf{Product mix shift}: The types of devices registered differ by jurisdiction.
\end{itemize}
We perform three targeted tests to assess these explanations (Section~\ref{sec:class1_tests}). The results strengthen the residual-versus-positional interpretation but do not fully exclude the alternatives.

Perturbation-based reclassification analysis (Supplementary Figure~S8) characterises the classifier's decision boundaries: reclassification succeeds in 100.0\% of NMPA cases, 93.0\% of FDA cases, and 84.2\% of EU~MDR cases, with Class~I to Class~III transitions the hardest (68.3\% success).

\subsection{Robustness and Sensitivity Analyses}

\subsubsection{Robustness Checks}
\label{sec:leakage_audit}
\label{sec:model_robustness}

Grouped cross-validation does not change the NMPA or FDA cross-jurisdictional transfer results. Exact text duplicates account for 11.9\% of NMPA records, 6.9\% of FDA records, and 4.9\% of EU~MDR records, yet $\Delta$F1 remains at +0.023 for NMPA and +0.006 for FDA (Supplementary Table~S6). The same pattern also appears across all three model families (Random Forest, Logistic Regression, and Gradient Boosting) with a modest contribution for NMPA and a negligible one for FDA (Supplementary Table~S7). This suggests that the NMPA result is not specific to Random Forest.

\subsubsection{Class~I Zero-Shot Failure: Mechanism Tests}
\label{sec:class1_tests}

To adjudicate among the competing explanations for the EU~MDR Class~I zero-shot collapse, we conduct three targeted tests.

\paragraph{Test A: Rebalancing class priors.} We undersample EU~MDR Class~I from 9,819 to 325 devices to match the $\sim$2.5\% Class~I prevalence in FDA and NMPA. Under these matched priors, Class~I F1 remains effectively zero (0.000), ruling out class-prior mismatch as the primary explanation.

\paragraph{Test B: Excluding Class~I.} Evaluating zero-shot transfer on EU~MDR Class~II and III only (excluding Class~I) yields a macro-F1 of 0.745 (Class~II F1 = 0.969, Class~III F1 = 0.522). This indicates that cross-jurisdictional transfer for higher-risk devices is substantially more successful than the overall macro-F1 of 0.384 suggests, and that the Class~I failure is the dominant source of degradation.

\paragraph{Test C: Error analysis.} Of the 9,819 EU~MDR Class~I devices, 97.9\% (9,615) are predicted as Class~II under zero-shot conditions. The factor profile of these devices shows that 98.1\% have \texttt{reusable}=1 and near-zero values on all other factors, confirming that EU~MDR Class~I is defined by the \textit{absence} of risk triggers. Models trained on FDA and NMPA, where Class~I devices have positionally assigned characteristics, lack a representation for this residual category.

These tests strengthen the residual-versus-positional interpretation as the most plausible explanation but cannot fully exclude contributions from text domain shift or extraction asymmetry.

\subsubsection{Unified Model: Effect of Factors}
\label{sec:unified_ablation}

Removing regulatory factors from the unified model reduces EU~MDR F1-macro from 0.984 to 0.785, confirming that high EU~MDR performance is driven by circular factor derivation and should not be cited as evidence of cross-jurisdictional compatibility. NMPA and FDA effects are modest (+0.031 and +0.017 respectively; Supplementary Table~S8).

\subsubsection{Reverse-Direction Transfer Probes}
\label{sec:reverse_transfer}

The main analysis examines transfer in one direction only, using EU~MDR-derived factors to classify NMPA and FDA devices. To assess whether this pattern depends on factor source, we construct two additional factor sets from the other jurisdictions and apply them to the remaining systems.

For NMPA, we derive seven features from the catalogue prefix structure, including prefixes 03 and 13 for implants and 68 for IVDs. For non-NMPA devices, these features are extracted by keyword matching. For FDA, we derive seven features from product-code panel families, including cardiovascular, orthopaedic, and IVD clusters. For non-FDA devices, these features are likewise extracted by keyword matching.

\begin{table}[t]
\centering
\caption{Reverse-direction transfer matrix: $\Delta$F1 by factor source and target jurisdiction (jurisdiction-specific pipelines). Parenthesised values: circular ceiling cases. Bold: only non-trivial cross-jurisdictional transfer.}
\label{tab:reverse_transfer}
\begin{tabular}{lccc}
\toprule
\textbf{Factor source} & \textbf{Target: EU~MDR} & \textbf{Target: NMPA} & \textbf{Target: FDA} \\
\midrule
EU~MDR factors & (+0.189) & \textbf{+0.023} & +0.007 \\
NMPA factors & +0.004 & (+0.021) & +0.005 \\
FDA factors & $-$0.001 & +0.000 & (+0.008) \\
\bottomrule
\end{tabular}
\end{table}

Table~\ref{tab:reverse_transfer} shows that transfer is directional rather than symmetric. EU~MDR factors improve NMPA classification, but this effect does not reverse, as NMPA-derived factors add little when applied to EU~MDR. FDA-derived factors likewise contribute almost no signal outside the source jurisdiction. The only non-trivial cross-jurisdictional result in the matrix is therefore the EU~MDR-to-NMPA transfer.

This pattern suggests that the shared signal between EU~MDR and NMPA is limited and one-way. It also suggests that FDA factor structure, which is tied to predicate-based classification, does not travel to the other systems in any meaningful way. The reverse-direction probes therefore narrow the interpretation of the main result. What appears is not broad compatibility across regimes, but a specific directional overlap between EU~MDR and NMPA.

\subsubsection{Text Encoder Sensitivity}
\label{sec:multilingual}

The main ablation uses a TF-IDF text baseline built from character n-grams. Because this representation produces largely disjoint feature spaces across Chinese and Latin scripts, the observed factor gain could in principle be inflated by a weak text channel rather than by genuine regulatory signal. We therefore re-run the analysis under two alternative specifications, using the same jurisdiction-specific factor extraction and cross-validation protocol as in the main ablation (Table~\ref{tab:master_results}), which is Random Forest with \texttt{random\_state=42}, five-fold stratified cross-validation, and seed = 42.

\paragraph{SVD component sweep.}
The NMPA factor gain varies substantially with the dimensionality of the TF-IDF representation, ranging from +0.003 to +0.026. The largest gain appears at 10 SVD components, which is also the setting in which the text baseline is weakest. This pattern suggests that the estimated factor contribution under TF-IDF is sensitive to text encoding capacity rather than fixed across specifications (Supplementary Figure~S8; Table~S9).

\paragraph{Multilingual sentence embeddings and device-type control.}
We next replace TF-IDF with 384-dimensional multilingual MiniLM embeddings \cite{reimers2019sentence} and cross this with a device-type control using 39 one-hot category indicators. This yields a two-by-two design that separates two possible explanations for the NMPA result: (i) weak text encoding and (ii) device-category confounding.

\begin{figure}[tbp] 
\centering \includegraphics[width=\columnwidth]{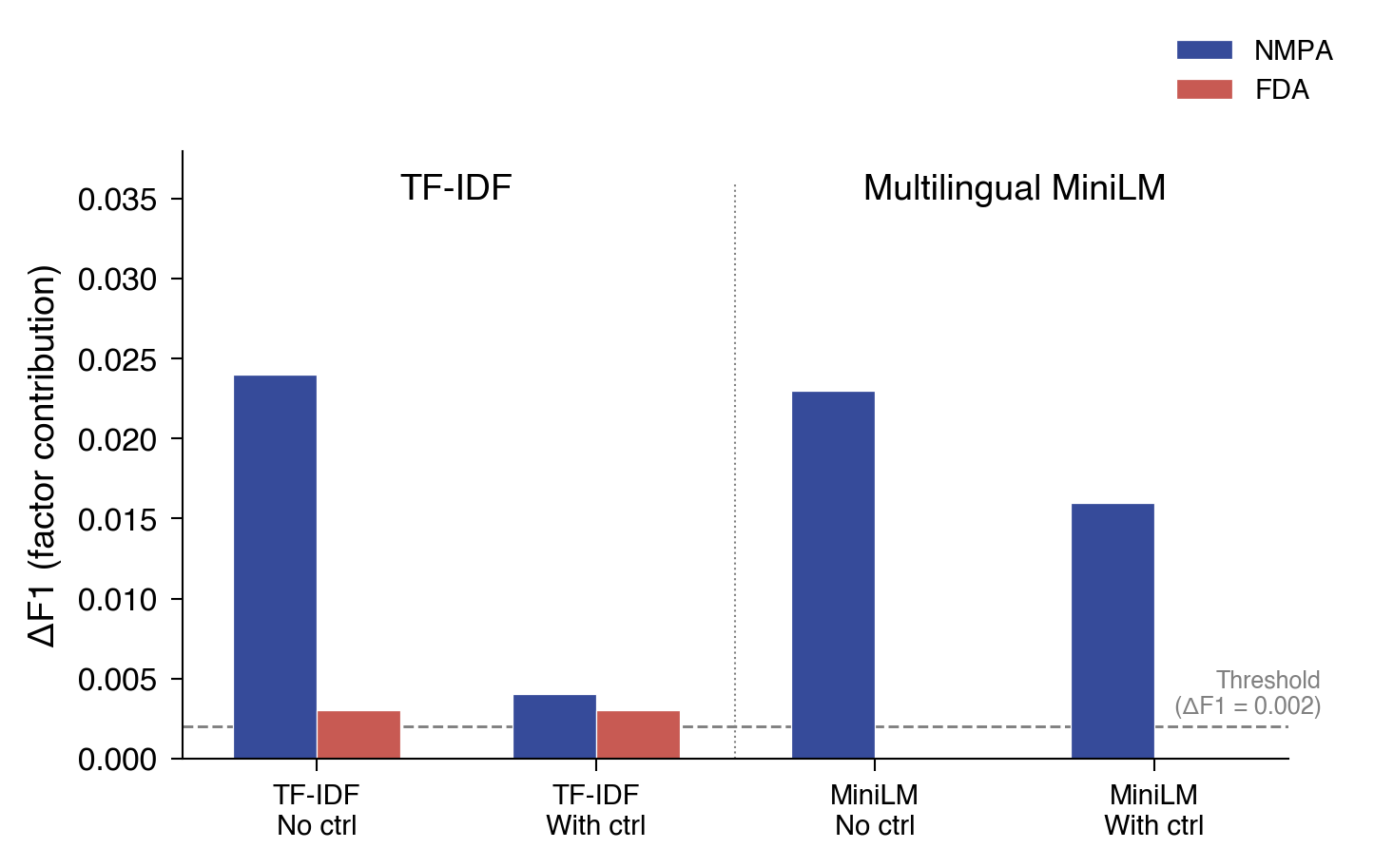} \caption{Two-by-two adjudication of NMPA factor contribution ($\Delta$F1), crossing text encoder (TF-IDF vs.\ multilingual MiniLM) with device-type control (39 one-hot NMPA category codes). NMPA factor gain is positive in all four conditions, but smaller once device-type control is added. Under TF-IDF, $\Delta$F1 falls from +0.024 to +0.004 with control; under MiniLM, it falls from +0.023 to +0.016. FDA values remain negligible in all conditions ($\leq +0.003$). The figure therefore indicates that the NMPA signal is reduced, but not eliminated, by device-type control, and that this residual is larger under the stronger multilingual encoder.} 
\label{fig:adjudication} 
\end{figure}

The first comparison is across encoders without device-type control. Here the estimated NMPA factor gain is nearly unchanged, moving from +0.024 under TF-IDF to +0.023 under MiniLM. The second comparison adds device-type control. Under this stricter specification, the factor gain falls, but it does not disappear: $\Delta$F1$_{\mathrm{ctrl}}$ is +0.004 under TF-IDF and +0.016 under MiniLM. In other words, strengthening the text representation does not remove the NMPA signal, and controlling for device category reduces the gain but does not eliminate it. Figure~\ref{fig:adjudication} visualises this pattern. The key point is that the residual is larger under the stronger multilingual encoder than under TF-IDF, which is difficult to reconcile with the view that the original NMPA gain was only an artefact of weak text features.

The same multilingual specification also improves zero-shot transfer, raising macro-F1 from 0.431 to 0.494, with the largest improvement in Class~III. Class~I, however, still collapses to below 0.03. This matters because it narrows the interpretation of the failure. Better cross-lingual text representations help overall transfer, but they do not resolve the Class~I breakdown. That failure therefore appears to arise from structural mismatch rather than from language alone. Supplementary Table~S10 reports the full multilingual sensitivity results.

\section{Discussion}
\label{sec:discussion}

\subsection{Factor Transfer: A Fragile Signal}

The central question is whether EU~MDR-derived risk factors carry cross-jurisdictional classification signal. The answer depends critically on how the comparison is constructed.

Under the symmetric extraction pipeline (Table~\ref{tab:master_results}), which provides the epistemically cleanest comparison, factor contribution is near zero for all three jurisdictions ($\Delta$F1 $<$ 0.01). Without pipeline-specific advantages, the cross-jurisdictional signal is at best marginal and likely indistinguishable from noise.

In jurisdiction-specific extraction pipelines, the NMPA gain ($\Delta$F1 = +0.024) is modest but reproducible. These are stable under grouped cross-validation (Section~\ref{sec:leakage_audit}), directionally consistent across three model families (Section~\ref{sec:model_robustness}), and the only cross-jurisdiction link where the 95\% CI (computed from 10 seed-level means) excludes zero. However, this gain must be interpreted as an upper bound that conflates genuine concept transfer with pipeline asymmetry, text encoding limitations, and device-category confounding. The six-fold ratio between the jurisdiction-specific and symmetric estimates ($+0.024$ vs.\ $+0.004$) indicates that at least 80\% of the observed effect is attributable to pipeline design rather than cross-jurisdictional concept overlap.

Sensitivity analyses resolve this ambiguity through a complete two-by-two design crossing text encoder (TF-IDF vs.\ multilingual MiniLM) with device-type control (absent vs.\ present). Under TF-IDF with device-type controls, the factor contribution survives but attenuates substantially ($+$0.004; Figure~\ref{fig:adjudication}), indicating partial confounding with device-category information. Under multilingual embeddings with the same device-type controls, the contribution \emph{survives} more robustly ($+$0.016), confirming genuine regulatory signal beyond device-category structure.

\begin{enumerate}
\item Under both text encoders, the factor signal survives device-type control, ruling out pure device-category confounding as the explanation for the NMPA gain.
\item The larger residual under multilingual embeddings ($+$0.016 vs.\ $+$0.004) confirms that stronger text encoding reveals additional regulatory structure not accessible to TF-IDF.
\end{enumerate}
The practical implication is that  the magnitude of the factor contribution depends on the text encoding context. TF-IDF-based estimates are conservative lower bounds; multilingual embeddings provide a more informative characterisation of genuine cross-jurisdictional regulatory signal.

The FDA result ($\Delta$F1 = +0.006 jurisdiction-specific, +0.003 symmetric) is negligible and consistent across all analyses. Reverse-direction probes confirm that FDA-derived factors produce no meaningful transfer in any direction (Section~\ref{sec:reverse_transfer}), suggesting that predicate-based classification operates through an information structure that is fundamentally opaque to factor-based probes.

\subsection{Shared Risk Intuitions}

The concept-to-class association analysis reveals that the implantable and invasiveness factors are associated with the highest risk class across all three jurisdictions despite their different classification mechanisms. This qualitative pattern, namely that implantable devices are assigned to the highest risk class regardless of whether the jurisdiction classifies through explicit rules, predicate chains, or catalogue matching, is robust across all three systems. These shared risk intuitions represent a potential foundation for cross-jurisdictional regulatory harmonisation efforts, even in the absence of procedural alignment. We note that the Cram\'{e}r's $V$ magnitudes are not directly comparable across jurisdictions due to differing class distributions (EU~MDR has 1.9\% Class~III devices versus approximately 40--45\% for FDA and NMPA), and the qualitative finding of shared directionality should be emphasised over the specific numerical values.

\subsection{Failure Structure and Cross-Lingual Limitations}

The zero-shot transfer result, in which Class~I F1 falls to 0.001, indicates a severe failure when applying models trained on FDA and NMPA data to EU~MDR. Multiple mechanisms contribute to this failure, and the study design cannot fully disentangle them.

A previously unacknowledged factor is the cross-lingual text barrier. The TF-IDF text features are computed on mixed Chinese (NMPA), English (FDA), and multilingual European (EU~MDR) text using character n-grams. Because Chinese and Latin character vocabularies are almost entirely disjoint, the resulting 42-dimensional SVD features effectively partition by language rather than capturing shared semantic content. In the unified model and zero-shot experiments, the text channel is therefore largely non-functional across jurisdictions, meaning the model relies almost entirely on the seven regulatory factors for cross-jurisdictional signal. This interpretation is supported by the multilingual embedding analysis (Section~\ref{sec:multilingual}), which shows that replacing TF-IDF with cross-lingually aligned sentence embeddings substantially changes the zero-shot transfer results.

The Class~I collapse specifically is most parsimoniously explained by the conjunction of two factors: (i) the text features carry no cross-lingual information, forcing the model to rely on regulatory factors; and (ii) EU~MDR Class~I is defined residually (absence of risk triggers), which the seven binary/ordinal factors cannot represent because ``absence of all factors'' is also the default for undeterminable devices.

The native four-class evaluation confirms that class compression contributes negligibly to this failure, with macro-F1 essentially identical under both the compressed three-class (0.804) and native four-class (0.804) schemes (Supplementary Table~S4).

\subsection{Pipeline Asymmetry and Epistemic Status}

A methodological consideration that warrants explicit discussion is the asymmetry among the three factor extraction pipelines. EU~MDR factors are derived primarily from the recorded Annex~VIII classification rule, supplemented by multilingual keyword matching. NMPA factors are inferred from the two-digit classification code prefix plus keywords. FDA factors are extracted from text via keyword matching alone. These pipelines differ in three respects. These include proximity to the classification label, degree of source structure, and the amount of human-encoded regulatory logic they contain. Specifically, the EU~MDR pipeline is closest to the label (since rules directly determine the class), the NMPA pipeline occupies an intermediate position (code prefixes encode broad device categories but not classification logic per se), and the FDA pipeline is furthest from the label.

This asymmetry is a design-defining limitation, not a resolved objection. It means that the study supports two distinct estimands:
\begin{enumerate}
\item \textbf{Primary comparative estimand}: what remains under a symmetric extraction regime. This isolates cross-jurisdictional structure from pipeline artefacts and is the epistemically appropriate basis for claims about regulatory concept overlap.
\item \textbf{Applied upper bound}: what one could achieve with the best available jurisdiction-specific extraction pipeline. This answers the practical question but conflates concept transfer with pipeline design.
\end{enumerate}

The symmetric pipeline (Section~\ref{sec:symmetric_pipeline}) yields near-zero factor contribution for all jurisdictions (all $\Delta$F1 $<$ 0.01), indicating that the clean cross-jurisdictional signal is marginal. Manual annotation (Section~\ref{sec:manual_annotation}) quantifies extraction fidelity but does not resolve whether the jurisdiction-specific gains reflect genuine concept overlap or extraction artefacts.

\subsection{Limitations}

The EU~MDR factor derivation introduces circularity because factors are extracted from the classification rule recorded in the database. This contaminates the EU~MDR result, which must be treated as a source-jurisdiction ceiling rather than a commensurable data point alongside NMPA and FDA.

Pipeline asymmetry across the three extraction methods is a design-defining limitation. The manual annotation and symmetric pipeline analyses mitigate but do not close this concern, and the possibility remains that differential extraction fidelity contributes to the observed pattern between NMPA and FDA.

Although we now test three factor sources (EU~MDR, NMPA, and FDA-derived), each factor set represents one operationalisation of that jurisdiction's classification logic. Alternative operationalisations may yield different results. The reverse-direction probes (Section~\ref{sec:reverse_transfer}) confirm that transfer is source-dependent and asymmetric, which limits the scope of generalisable conclusions about inter-jurisdictional compatibility.

The text baseline (TF-IDF with character n-grams, reduced to 42 SVD components) is architecturally incapable of cross-lingual transfer. In the per-jurisdiction experiments, this is benign because each jurisdiction operates within a single language domain. In the unified and zero-shot experiments, however, the text features carry near-zero cross-lingual information, meaning the cross-jurisdictional results are effectively testing factor-only transfer with 42 dimensions of language-indicator noise. Multilingual sentence embeddings partially address this limitation (Section~\ref{sec:multilingual}), but the text baseline should be considered a lower bound on text representational capacity.

EUDAMED records contain limited textual information, averaging only 36 characters per product description. This text poverty constrains the factor extraction quality for EU~MDR devices and means that the EU~MDR text-only analysis is structurally weak. Under grouped cross-validation, EU~MDR text-only F1 drops substantially, confirming that near-duplicate memorisation inflates text-only performance for this jurisdiction.

The seven regulatory factors are correlated with device product categories (e.g., ``implantable'' correlates with orthopaedic devices). The two-by-two device-type control analysis (Section~\ref{sec:multilingual}) shows that the NMPA factor signal survives under both encoders (MiniLM: $+$0.016; TF-IDF: $+$0.004), providing evidence that the signal is not purely a device-type artefact. However, the 39 category codes used as controls are derived from NMPA classification prefixes, introducing partial circularity: the survival of factors beyond these categories does not exclude confounding with finer-grained device-type information not captured by the 39-category encoding. Additionally, EU~MDR device categories are uniformly ``Unknown'' in EUDAMED, so the device-type control is uninformative for this jurisdiction.

Manual validation uses 150 devices per jurisdiction with inter-rater reliability assessed on a 90-device subsample after a structured adjudication round (mean $\kappa$ = 0.781, substantial agreement). Active device exhibited the lowest $\kappa$ (0.577) despite 89.7\% raw agreement, reflecting $\kappa$ attenuation under skewed prevalence. The adjudication process, while necessary to resolve codebook ambiguities (particularly for invasiveness), means that the reported $\kappa$ values represent post-calibration reliability rather than na\"{\i}ve agreement. The 51.7\% undeterminable rate for EU~MDR reflects the extreme text brevity and limits the weight this validation can bear.

Ground truth classification labels are derived from database records that capture regulatory outcomes rather than the reasoning processes that produced those outcomes.

\section{Conclusion}

This paper examined whether risk concepts derived from EU~MDR Annex~VIII provide portable classification signal across three major regulatory systems. Using AI classification as an empirical probe, we tested whether seven EU~MDR-derived factors improved device classification in EU~MDR, NMPA, and FDA data.

The primary result is negative. Under the symmetric extraction pipeline, which provides the cleanest cross-jurisdictional comparison, factor contribution is near zero in all three jurisdictions ($\Delta$F1 $<$ 0.01). A non-trivial result appears only under jurisdiction-specific pipelines, where EU~MDR-derived factors yield a modest gain in the NMPA setting. That gain remains positive under stronger multilingual text encoding and device-type control, but it is still limited in magnitude and should be interpreted as an upper-bound estimate rather than as evidence of general cross-jurisdictional transfer. Reverse-direction probes further show that the effect is directional: NMPA-derived factors do not transfer back to EU~MDR, and FDA-derived factors add almost no signal outside the source jurisdiction.

These findings have two implications for explainable regulatory AI. First, shared regulatory vocabulary should not be treated as evidence that classification logic is portable across jurisdictions. Cross-jurisdictional models need explicit empirical validation rather than transfer assumptions. Second, estimates of factor utility depend materially on how concepts are extracted and how text is represented. In this setting, the text encoder is not a neutral implementation detail; it changes the measured contribution of regulatory factors.

More broadly, the study shows that concept-based XAI can be used not only to interpret model behaviour, but also to test the limits of model transfer across regulatory environments. For medical device informatics, this provides a practical framework for evaluating when regulatory concepts support cross-jurisdictional modelling and when they do not. The results suggest that transfer is limited, directional, and method-sensitive, and that robust deployment claims should therefore be based on measured portability rather than on surface similarity between regulatory systems.

\section*{Code Availability}

The implementation used in this study is publicly available at:
\url{https://github.com/RegAItool/explain_reg_AI}

\bibliographystyle{IEEEtran}
\bibliography{medical_device_references}

\end{document}